\documentclass[copyright,creativecommons]{eptcs}
\providecommand{\event}{QPL 2013} 
\usepackage{breakurl,amsmath,amsthm,amssymb,pb-diagram}             

\title{Duality Theory and Categorical Universal Logic:\\
With Emphasis on Quantum Structures}
\author{Yoshihiro Maruyama
\institute{Quantum Group\\ Department of Computer Science\\ University of Oxford}
\email{maruyama@cs.ox.ac.uk}
}
\def\titlerunning{Duality Theory and Categorical Universal Logic}
\def\authorrunning{Y. Maruyama}

\providecommand{\bibitemdeclare}[2]{}
\providecommand{\surnamestart}{}
\providecommand{\surnameend}{}
\providecommand{\urlprefix}{Available at }
\providecommand{\url}[1]{\texttt{#1}}
\providecommand{\href}[2]{\texttt{#2}}
\providecommand{\urlalt}[2]{\href{#1}{#2}}
\providecommand{\doi}[1]{doi:\urlalt{http://dx.doi.org/#1}{#1}}
\providecommand{\bibinfo}[2]{#2}

\newtheorem{thm}{Theorem}[section]

\newtheorem{cor}[thm]{Corollary}
\newtheorem{prop}[thm]{Proposition}
\newtheorem{defi}[thm]{Definition}

\begin{document}
\maketitle

\begin{abstract}
Categorical Universal Logic is a theory of monad-relativised hyperdoctrines (or fibred universal algebras),
which in particular encompasses categorical forms of both first-order and higher-order quantum logics
as well as classical, intuitionistic, and diverse substructural logics. Here we show
there are those dual adjunctions that have inherent hyperdoctrine structures in their predicate functor parts.
We systematically investigate into the categorical logics of dual adjunctions 
by utilising Johnstone-Dimov-Tholen's duality-theoretic framework.
Our set-theoretical duality-based hyperdoctrines for quantum logic have both universal and existential quantifiers
(and higher-order structures), 
giving rise to a universe of Takeuti-Ozawa's quantum sets via the tripos-to-topos construction by Hyland-Johnstone-Pitts.
The set-theoretical hyperdoctrinal models of quantum logic, as well as all quantum hyperdoctrines with cartesian base categories,
turn out to give sound and complete semantics for Faggian-Sambin's first-order quantum sequent calculus over cartesian type theory;
in addition, quantum hyperdoctrines with monoidal base categories are sound and complete for the calculus over linear type theory.
We finally consider how to reconcile Birkhoff-von Neumann's quantum logic and 
Abramsky-Coecke's categorical quantum mechanics (which is modernised quantum logic as an antithesis to the traditional one) via categorical universal logic.
\end{abstract}

\section{Introduction}

Different sorts of categorical logic have been developed in the last few decades, 
including categorical intuitionistic logic (see, e.g., Johnstone \cite{Jo2}) and
categorical quantum logic (see, e.g., Heunen-Jacobs \cite{HJ}; Jacobs \cite{Jacnew}). 
However, a unifying perspective upon various categorical logics is still lacking, 
which is the ultimate aim of this work, and towards which we take a first step in the present paper. 

To this end, we rely upon a monad-relativised concept of Lawvere's hyperdoctrine \cite{Law1};
the reason is as follows. Let us consider how we can unify, e.g., 
toposes and dagger kernel categories (in the sense of Heunen-Jacobs \cite{HJ}).
Although they appear to be rather different as single categories, nevertheless, 
the logical functorical substances of them are not so different: 
a topos ${\bf E}$ induces the subobject functor 
$${\rm Sub}_{\bf E}(\mbox{-}):{\bf E}^{\rm op}\to {\bf HA}$$
where ${\bf HA}$ is the category of heyting algebras
(there is an adjunction between toposes and higher-order hyperdoctrines; see Frey \cite{Frey});
a dagger kernel category ${\bf H}$ induces the kernel subobject functor
$${\rm KSub}_{\bf H}(\mbox{-}):{\bf H}^{\rm op}\to {\bf OML}$$
where ${\bf OML}$ is the category of orthomodular lattices
(subtleties on morphisms do not matter here). 
What is essential in interpreting logical concepts (e.g., quantifiers) is 
this fibrational or hyperdoctrinal structure, as is well known (see, e.g., Jacobs \cite{Jac}).

We thus define a monad-relativised hyperdoctrine as a functor (or algebra-valued presheaf)
$$P:{\bf C}^{\rm op}\to {\bf Alg}(T)$$
with suitable conditions to express logical concepts
where $T$ is a monad on ${\bf Set}$, 
which amounts to a (possibly infinitary) variety in terms of universal algebra.
We call our theory of monad-relativised hyperdoctrines (or fibred universal algebras) Categorical Universal Logic.
Choosing different monads or varieties, we can treat different sorts of categorical logic.
For instance, Maruyama \cite{Mar} shows that any axiomatic extension of the non-commutative Full Lambek calculus 
(see, e.g., Galatos et al. \cite{GJKO}),
which encompasses classical, intuitionistic, linear, fuzzy, and relevant logics, can be given 
sound and complete semantics via the corresponding class of monad-relativised hyperdoctrine.

In the present paper, we show that 
this is even true in the case of Faggian-Sambin's first-order quantum sequent calculus \cite{FS},
which has both universal and existential quantifiers, 
moreover enjoying excellent proof-theoretic properties such as cut elimination.
We consider the calculus over either of cartesian and monoidal type theory, 
or quantum hyperdoctrines with either of cartesian and monoidal base categories.
Note that Heunen-Jacobs \cite{HJ} discusses quantified quantum logic, but does not give a completeness result 
with respect to any proof-theoretic calculus, and does not treat universal quantifier in an adequate manner 
(indeed, they prove that universal quantifier only exists in boolean dagger-kernel categories, 
whereas existential quantifier always exists). Note also that 
Faggian-Sambin's calculus can be adapted so as to express features of quantum physics and information, 
such as entanglement (see, e.g., Zizzi \cite{Zizzi} and Battilotti-Zizzi \cite{BZ}).

A general question is how we can construct models of monad-relativised hyperdoctrines.
We consider duality does the job; in this paper, a duality means a dual adjunction. 
Let us think of the well-known dual adjunction between frames ${\bf Frm}$ and topological spaces ${\bf Top}$.
Frames give the propositional logic of open sets. The predicate functor of the dual adjunction 
$${\mathcal O}:{\bf Top}^{\rm op} \to {\bf Frm}$$
then turns out to have existential quantifier (in Lawvere's sense). 
Note that topological geometric logic (i.e., the quantified logic of open sets) does not have universal quantifier,
since open sets are not necessarily closed under arbitrary intersections.
We thus think that duality for propositional logic is a hyperdoctrinal model of predicate logic.

In order to discuss such phenomena in a systematic way, 
we use Johnstone-Dimov-Tholen's duality-theoretic framework (the main idea is due to Johnstone's ``general concrete dualities" \cite[VI.4]{Jo2}; however, certain technical points have only been explicated later, by Dimov-Tholen \cite{DT} and its expository companion Porst-Tholen \cite{PT}; some details are explained in Maruyama \cite{Marx}).
They basically think of two concrete categories ${\bf C}$ and ${\bf D}$
(concreteness means the existence of faithful functors into ${\bf Set}$), and 
assume $\Omega$ living in both ${\bf C}$ and ${\bf D}$, finally 
${\rm Hom}_{\bf C}(\mbox{-},\Omega)$ and ${\rm Hom}_{\bf D}(\mbox{-},\Omega)$ 
yielding a dual adjunction between ${\bf C}$ and ${\bf D}$.
In our case, one of ${\bf C}$ and ${\bf D}$, say ${\bf D}$, is ${\bf Alg}(T)$.
Based upon this general setting, we consider when the predicate functor 
$${\rm Hom}_{\bf C}(\mbox{-},\Omega):{\bf C}^{\rm op}\to {\bf Alg}(T)$$
of such a dual adjunction has a hyperdoctrine structure.
We give general criteria, and apply them to concrete situations including 
dual adjunctions for convex and quantum structures as well as the topological one mentioned above
(note that ${\mathcal O}$ above may be seen as ${\rm Hom}_{\bf Top}(\mbox{-},{\bf 2})$).
If the base category ${\bf C}$ of the predicate functor is ${\bf Set}$, 
then ${\rm Hom}_{\bf C}(\mbox{-},\Omega)$ always has both universal and existential quantifiers 
(and higher-order structures as well). 

In particular, we look at the case that $\Omega$ is the lattice of projection operators 
(or closed subspaces) on a Hilbert space. In this case, the set-based quantum hyperdoctrine 
${\rm Hom}_{\bf Set}(\mbox{-},\Omega)$ gives rise to a universe of Takeuti-Ozawa's quantum-valued sets
(see Takeuti \cite{Tak} and Ozawa \cite{Oza}) via the tripos-to-topos construction,
which is originally due to Hyland-Johnstone-Pitts \cite{HJP}.
We can then refine the completeness result for Faggian-Sambin's first-order quantum logic 
into that with respect to these set-based Tarskian models only (rather than all models).

The rest of the paper is organised as follows.
In Section \ref{mrhyp}, we introduce the concept of hyperdoctrines relativised to monads $T$.
In Section \ref{logdual}, we investigate into the categorical logics of dual adjunctions 
in a general setting based upon Johnstone-Dimov-Tholen's duality theory.
In Section \ref{geom}, we illustrate applications to convex and topological geometric logics,
constructing duality models of them.
In Section \ref{quant}, we discuss quantified quantum logic and quantum set theory
from our point of view, establishing hyperdoctrinal completeness results 
for Faggian-Sambin's quantum sequent calculus (over either of cartesian and monoidal type theory), 
and relating our set-theoretical hyperdoctrinal models to Takeuti-Ozawa's quantum-valued models of set theory.
We finally discuss how to reconcile Birkhoff-von Neumann's quantum logic and 
Abramsky-Coecke's categorical quantum mechanics via the idea of categorical universal logic.


\section{Monad-Relativized Hyperdoctrines}\label{mrhyp}

A hyperdoctrine comes with a base category ${\bf C}$ 
and a contravariantly functorial assignment $P$ of logical algebras to objects in ${\bf C}$.
Here, ${\bf C}$ represents a type theory or a structure of domains of discourse.
Given $C\in {\bf C}$, $P(C)$ represents an algebra of predicates or proposiitons on $C$,
and, for an arrow $f:C\to D$ in ${\bf C}$, 
$F(f)$ translates propositions on $D$ into those on $C$, 
and amounts to substitution from a syntactical point of view.

In the concept of hyperdoctrine, thus, types and propositions are not primarily supposed to 
be equivalent, in contrast to the Curry-Howard-Lambek isomorphism perspective.
Types are represented by one category, and propositions by another algebraic category.
The hyperdoctrinal methodology gives us more flexibility 
than the Curry-Howard-Lambek one, since 
the type structure and proposition structure of logic can be totally different in the concept of hyperdoctrine. 

Accordingly, we can freely combine type theory and logic by means of hyperdoctrines,
whereas, in the Curry-Howard-Lambek approach, 
type theory and logic must come in harmony from the very beginning; however,
there seems to be no reason for presupposing such a priori harmony between type theory and logic.
From such a point of view, we could say that logic and type theory should turn out to be equivalent 
after their independent births even if they are equivalent in the end.

We consider that this feature of hyperdoctrines is particularly significant 
in formulating the logic of quantum mechanics, 
for the logic of quantum propositions differs from the logic (or type theory) of quantum systems: 
the former is given by traditional quantum logic a la Birkhoff-von Neumann, 
and the latter by categorical quantum mechanics a la Abramsky-Coecke \cite{AC}.

In the following, $T$ denotes a monad, and in order to enable Lawvere's definition of quantifiers as adjoints,
we assume that each $T$-algebra $A$ is equipped with a partial order $\leq$ preserved under homomorphisms, 
which we call the deducibility ordering of $A$. Rather than directly assuming this, 
we may alternatively assume that the finitary powerset monad is a submonad of $T$, so that 
deducibility orderings are derived, i.e.,
each $T$-algebra has a semilattice reduct with an intrinsic partial order,
which is automatically preserved by homomorphisms of $T$-algebras.

\begin{defi}
A $T$-hyperdoctrine (or fibred $T$-algebra) is defined 
as an ${\bf Alg}(T)$-valued presheaf 
$$P:{\bf C}^{\rm op}\to {\bf Alg}(T)$$
where ${\bf C}$ is a 
category with finite products. 
For an arrow $f$ in ${\bf C}$,
$P(f)$ is called the pullback of $f$.
For $C\in {\bf C}$, $P(C)$ is called the fibre of $P$ over $C$.

We then define the following notions:
\begin{itemize}
\item A $T$-hyperdoctrine $P:{\bf C}^{\rm op}\to {\bf Alg}(T)$ has universal quantifier $\forall$ iff,
for any projection $\pi:X\times Y\to Y$ in ${\bf C}$, the following functor 
$$P(\pi):P(Y)\to P(X\times Y)$$
has a right adjoint, denoted 
$$\forall_{\pi}:P(X\times Y)\to P(Y)$$
and the corresponding Beck-Chevalley condition holds, i.e., the following diagram commutes
for any arrow $f:Z\to Y$ in ${\bf C}$ ($\pi':X\times Z\to Z$ below denotes a projection):
$$
\begin{diagram}             
\node{P(X\times Y)} \arrow{s,l}{P(X\times f)} \arrow{e,l}{\forall_{\pi}} \node{P(Y)} \arrow{s,r}{P(f)}\\
\node{P(X\times Z)} \arrow{e,r}{\forall_{\pi'}} \node{P(Z)}
\end{diagram}
$$
Note that $P(X)$ and the like above are equipped with partial orders, 
thanks to our assumption mentioned above.
Note that $P(X)$ and the like above are seen as categories; here
we are using the ``logicality of monad" assumption:
$T$-algebras come with ``deducibility" relations,
which yield categorical structures on $T$-algebras.
\item A $T$-hyperdoctrine $P:{\bf C}^{\rm op}\to {\bf Alg}(T)$ has existential quantifier $\exists$ iff,
for any projection $\pi:X\times Y\to Y$ in ${\bf C}$, $P(\pi):P(Y)\to P(X\times Y)$
has a left adjoint, which shall be denoted as 
$$\exists_{\pi}:P(X\times Y)\to P(Y)$$
and the corresponding Beck-Chevalley diagram commutes for any $f:Z\to Y$ in ${\bf C}$
($\pi':X\times Z\to Z$ below is a projection):
$$
\begin{diagram}             
\node{P(X\times Y)} \arrow{s,l}{P(X\times f)} \arrow{e,l}{\exists_{\pi}} \node{P(Y)} \arrow{s,r}{P(f)}\\
\node{P(X\times Z)} \arrow{e,r}{\exists_{\pi'}} \node{P(Z)}
\end{diagram}
$$
\item A $T$-hyperdoctrine $P:{\bf C}^{\rm op}\to {\bf Alg}(T)$ has equality $=$ iff,
for any diagonal $\delta:X\to X\times X$ in ${\bf C}$, the following functor
$$P(\delta):P(X\times X)\to P(X)$$
has a left adjoint, which shall be denoted as 
$${\rm Eq}_{\delta}:P(X)\to P(X\times X).$$
\end{itemize}
A quantified $T$-hyperdoctrine is defined as a $T$-hyperdoctrine having $\forall$ and $\exists$.
A first-order $T$-hyperdoctrine is defined as a $T$-hyperdoctrine having $\forall$, $\exists$, and $=$.
\end{defi}

In standard, categorical developments of regular, coherent, and intuitionistic logics,
Frobenius Reciprocity is usually assumed (or holds) as well as Beck-Chevalley conditions.
In the present paper, however, we do not generally assume Frobenius Reciprocity.

The main reason is that Frobenius Reciprocity is not appropriate for 
certain logical systems, including quantum logic;
recall that the Frobenius Reciprocity condition for existential quantifier involves the distributivity 
of $\exists$ over $\wedge$, a kind of infinitary distributivity law ($\exists$ may be seen as infinite joins),
which is not generally acceptable in quantum logic.

In sequent calculi with restricted context formuli or ``visibility", like Basic Logic by Sambin et al.,
Frobenius Reciprocity is actually harmful to obtain complete semantics.
Note that we always assume Beck-Chevalley conditions, since it is logically indispensable
to interpret the substitution of terms for variables. 
Categorical models without Beck-Chevalley properties are inadequate as semantics of logic,
however those without Frobenius Reciprocity are not necessarily so.

We sometimes consider quantified hyperdoctrines without one of $\exists$ and $\forall$.
For example, (topological) geometric logic only has existential quantifier, 
and thus it is natural to regard $T$-hyperdoctrines with $\exists$ only 
as being already quantified in the case of geometric logic.

The principle of comprehension in set theory can be understood in categorical terms of fibration
as originally discovered by Lawvere 
and Benabou. 
When we talk about comprehension, we assume that 
each $T$-algebra has a greatest element $\top$ with respect to its deducibility ordering,
and that greatest elements are preserved by homomorphisms of $T$-algebras.

$T$-hyperdoctrines can be seen as indexed categories.
We are therefore able to apply the Grothendieck construction to a $T$-hyperdoctrine 
$$P:{\bf C}^{\rm op}\to {\bf Alg}(T)$$
thus obtaining a fibred category $$\int P$$
which can be described as follows.
An object of $\int P$ is a pair $(X,a)$ where 
$X$ is an object of ${\bf C}$, and 
$a$ is an object of a $T$-algebra $P(X)$ seen as a category.
An arrow of $\int P$ from $(X,a)$ to $(Y,b)$ is a pair $(f,k)$ where 
$f$ is an arrow in ${\bf C}$ from $X$ to $Y$, 
and $k$ is an arrow in $P(X)$ from $a$ to $P(f)(b)$
(note that, in $P(X)$, at most one arrow exists between two objects).

\begin{defi}\label{compre}
A $T$-hyperdoctrine $P:{\bf C}^{\rm op}\to {\bf Alg}(T)$ has comprehension $\{ \mbox{-} \}$
iff the truth functor defined below
$$\top:{\bf C}\to \int P$$ 
has a right adjoint, which shall be denoted as
$$\{ \mbox{-} \}:\int P\to {\bf C}.$$ 
The truth functor $\top$ is defined as follows.
Concerning the object part, $\top$ maps an object $X\in {\bf C}$ to an object
$$(X,\top_{P(X)})\in \int P$$
where $\top_{P(X)}$ denotes the greatest element of a $T$-algebra $P(X)$.
Regarding the arrow part, $\top$ maps an arrow $f:X\to Y$ in ${\bf C}$ 
to an arrow $(f,!)$ in $\int P$ from $(X,\top_{P(X)})$ to $(Y,\top_{P(Y)})$
where $!$ is a unique arrow from $\top_{P(X)}$ to $P(f)(\top_{P(Y)})$, 
which equals $\top_{P(X)}$.
\end{defi}

Higher-order $T$-hyperdoctrines are defined by requiring 
additional conditions for higher type structures and object classifiers
as follows.

\begin{defi}
A higher-order $T$-hyperdoctrine (or a $T$-tripos) is defined as 
a $T$-hyperdoctrine $P:{\bf C}^{\rm op}\to {\bf Alg}(T)$ such that 
\begin{itemize}
\item the base category ${\bf C}$ is Cartesian closed;
\item $P$ has quantifiers $\forall$, $\exists$, equality $=$, and comprehension $\{\mbox{-}\}$;
\item $P$ has an object classifier in the following sense: there exists 
$$\Omega \in {\bf C}$$ 
such that $P$ is naturally equivalent to 
$${\rm Hom}_{\bf C}(\mbox{-},\Omega).$$
An object classifier is also called a generic object or truth value object. 
\end{itemize}
\end{defi}

If $T$ represents intuitionistic logic,
then higher-order $T$-hyperdoctrines (or $T$-triposes) 
basically amount to toposes,
for the following well-known fact (see, e.g., Jacobs \cite{Jac}).

\begin{prop}
A category ${\bf E}$ with pullbacks is a topos iff 
the induced subobject functor (whose arrow part is defined by taking pullbacks)
$${\rm Sub}_{\bf E}:{\bf E}^{\rm op}\to {\bf Pos}$$
is a higher-order $T$-hyperdoctrine for the monad $T$ 
whose algebras are categorically equivalent to Heyting algebras
(${\bf Pos}$ denotes the category of posets).
\end{prop}

We may thus consider that 
the concept of higher-order $T$-hyperdoctrines (or $T$-triposes)
logically correspond to the concept of toposes relativised to the monad $T$.

\section{Categorical Logic of Dual Adjunctions}\label{logdual}


Let us recall the setting of 
duality induced by schizophrenic objects
in the general style of Johnstone-Dimov-Tholen
(the term ``schizophrenic" may be inappropriate, 
but there is no widely accepted alternative).
That is, we have two categories ${\bf C}$ and ${\bf D}$ with
faithful functors $U:{\bf C}\to {\bf Set}$ and $V:{\bf D}\to {\bf Set}$,
and an object $\Omega$ which lives in both ${\bf C}$ and ${\bf D}$.
Then, two ${\rm Hom}$ functors 
${\rm Hom}_{\bf C}(\mbox{-},\Omega)$ and ${\rm Hom}_{\bf D}(\mbox{-},\Omega)$
give us a dual adjunction between ${\bf C}$ and ${\bf D}$ 
(under the assumption of initial lifting properties of $\Omega$; Maruyama \cite{Marx} gives a simpler account of the duality mechanism via what is called the harmony condition).

Now suppose that ${\bf D}$ is ${\bf Alg}(T)$, and ${\bf C}$ has finite products.
We are thus thinking of the following dual adjunction 
$${\rm Hom}_{{\bf Alg}(T)}(\mbox{-},\Omega) \dashv {\rm Hom}_{\bf C}(\mbox{-},\Omega):{\bf C}^{\rm op}\to {\bf Alg}(T).$$
Our proposal is to regard ${\rm Hom}_{\bf C}(\mbox{-},\Omega)$ as a $T$-hyperdoctrine.
We call $T$-hyperdoctrines arising in this way duality $T$-hyperdoctrines (or Stonean $T$-hyperdoctrines).
Note that the domain category ${\bf C}$ of a duality $T$-hyperdoctrine always comes with a faithful functor $U:{\bf C}\to {\bf Set}$.


According to our assumption,
every $T$-algebra is endowed with a partial order to represent a deductive relation.
In particular, $\Omega$ is thus endowed with a partial order $\leq_{\Omega}$,
which canonically induce a partial order on ${\rm Hom}_{\bf C}(X,\Omega)$ for $X\in {\bf C}$: i.e.,
$u \leq v$ for $u,v\in {\rm Hom}_{\bf C}(X,\Omega)$ iff for any $x\in U(X)$, 
$U(u)(x)\leq U(v)(x).$
In the following, we assume that $\Omega$ is complete with respect to the ordering $\leq_{\Omega}$.

When do duality $T$-hyperdoctrines have logical structures such as quantifiers?
The existence of adjoints of pullbacks of projections and diagonals can be shown
in quite general situations, 
as in the following propositions.
At the same time, however, Beck-Chevalley conditions are merely assumed in them.
Moreover they do not elucidate how those adjoints actually operate. 
Soon after the following three propositions, we prove more specialised propositions
in which Beck-Chevalley conditions are naturally accounted for, and then it becomes 
clearer how those adjoints representing logical constants operate.

\begin{prop}
Assume that a duality $T$-hyperdoctrine 
${\rm Hom}_{\bf C}(\mbox{-},\Omega):{\bf C}^{\rm op}\to {\bf Alg}(T)$
satisfies the following two conditions.
\begin{itemize}
\item For any $X\in {\bf C}$, ${\rm Hom}_{\bf C}(X,\Omega)$ has colimits (i.e., arbitrary joins).
\item The faithful functor $U:{\bf C}\to {\bf Set}$ associated with ${\bf C}$
commutes with colimits in the following sense: for any $X\in {\bf C}$ and 
any $f_i \in {\rm Hom}_{\bf C}(X,\Omega)$ where $i\in I$,
it holds that
$$\bigvee_{i\in I}(U(f_i))=U(\bigvee_{i\in I}f_i)$$
where $\bigvee_{i\in I}(U(f_i))$ is the meet of $\{ U(f_i) \ | \ i\in I\}$ in $\Omega^{U(X)}$, 
i.e., for any $x\in X$,
$$(\bigvee_{i\in I}U(f_i))(x) = \bigvee_{i\in I} (U(f_i)(x)).$$
\end{itemize}
Then, the duality $T$-hyperdoctrine 
${\rm Hom}_{\bf C}(\mbox{-},\Omega):{\bf C}^{\rm op}\to {\bf Alg}(T)$ 
has universal quantifier $\forall$, if the corresponding Beck-Chevalley condition holds.
\end{prop}


\begin{prop}
Assume that a duality $T$-hyperdoctrine 
${\rm Hom}_{\bf C}(\mbox{-},\Omega):{\bf C}^{\rm op}\to {\bf Alg}(T)$
satisfies the following two conditions.
\begin{itemize}
\item For any $X\in {\bf C}$, ${\rm Hom}_{\bf C}(X,\Omega)$ has limits (i.e., arbitrary meets).
\item The faithful functor $U:{\bf C}\to {\bf Set}$
commutes with limits in the following sense: for any $X\in {\bf C}$ and 
any $f_i \in {\rm Hom}_{\bf C}(X,\Omega)$ where $i\in I$,
it holds that
$$\bigwedge_{i\in I}(U(f_i))=U(\bigwedge_{i\in I}f_i)$$
where $\bigwedge_{i\in I}(U(f_i))$ is the meet of $\{ U(f_i) \ | \ i\in I\}$ in $\Omega^{U(X)}$.
\end{itemize}
Then, the duality $T$-hyperdoctrine has existential quantifier $\exists$,
if the corresponding Beck-Chevalley condition holds.
\end{prop}


In the following propositions, the Beck-Chevalley conditions are not assumed but derived,
and the structure of quantifiers is then more transparent.

In the following propositions, we use lifting conditions analogous to 
the initial lifting conditions in Johnstone-Dimov-Tholen's dual adjunction theorem.

\begin{prop}\label{forall}
Consider a duality $T$-hyperdoctrine 
${\rm Hom}_{\bf C}(\mbox{-},\Omega):{\bf C}^{\rm op}\to {\bf Alg}(T)$
such that the associated faithful functor $U:{\bf C}\to {\bf Set}$ preserves products.
Given a projection $\pi:X\times Y\to Y$ in ${\bf C}$ and 
$v\in {\rm Hom}_{\bf C}(X\times Y,\Omega)$, we define 
$$A ^{\pi}_v:U(Y)\to \Omega$$  
as follows: for $y\in U(Y)$,
$$A ^{\pi}_v(y):=\bigwedge \{ U(v)(x,y) \ | \ x\in U(X) \}.$$
If ``$A^{\pi}_v$ lifts to $\forall_{\pi}$", i.e., there is 
$\forall_{\pi}: {\rm Hom}_{\bf C}(X\times Y,\Omega)\to {\rm Hom}_{\bf C}(Y,\Omega)$
such that for any $v\in {\rm Hom}_{\bf C}(X\times Y,\Omega)$,
$$\forall_{\pi}(v)\in {\rm Hom}_{\bf C}(Y,\Omega) \mbox{ and } U(\forall_{\pi}(v))=A^{\pi}_v,$$
then the duality $T$-hyperdoctrine 
${\rm Hom}_{\bf C}(\mbox{-},\Omega):{\bf C}^{\rm op}\to {\bf Alg}(T)$ has universal quantifier $\forall$.
\end{prop}

The case of existential quantifier $\exists$ can be treated in a similar manner:

\begin{prop}\label{exists}
Consider a duality $T$-hyperdoctrine 
${\rm Hom}_{\bf C}(\mbox{-},\Omega):{\bf C}^{\rm op}\to {\bf Alg}(T)$
such that $U:{\bf C}\to {\bf Set}$ preserves products.
Given a projection $\pi:X\times Y\to Y$ in ${\bf C}$ and 
$v\in {\rm Hom}_{\bf C}(X\times Y,\Omega)$, we define 
$E ^{\pi}_v:U(Y)\to \Omega$
as follows: for $y\in U(Y)$,
$$E ^{\pi}_v(y):=\bigvee \{ U(v)(x,y) \ | \ x\in U(X) \}.$$
If  ``$E^{\pi}_v$ lifts to $\exists_{\pi}$", i.e., there is 
$\exists_{\pi}: {\rm Hom}_{\bf C}(X\times Y,\Omega)\to {\rm Hom}_{\bf C}(Y,\Omega)$
such that for any $v\in {\rm Hom}_{\bf C}(X\times Y,\Omega)$,
$$\exists_{\pi}(v)\in {\rm Hom}_{\bf C}(Y,\Omega) \mbox{ and } U(\exists_{\pi}(v))=A^{\pi}_v,$$
then the duality $T$-hyperdoctrine 
${\rm Hom}_{\bf C}(\mbox{-},\Omega):{\bf C}^{\rm op}\to {\bf Alg}(T)$ has existential quantifier $\exists$.
\end{prop}

In the following case of equality, we explicitly use the least element of $\Omega$.

\begin{prop}\label{equality}
Consider a duality $T$-hyperdoctrine 
${\rm Hom}_{\bf C}(\mbox{-},\Omega):{\bf C}^{\rm op}\to {\bf Alg}(T)$
such that $U:{\bf C}\to {\bf Set}$ preserves products.
Given a diagonal $\delta:X\to X\times X$ in ${\bf C}$ and 
$v\in {\rm Hom}_{\bf C}(X,\Omega)$, we define 
$I^{\delta}_v:U(X\times X)\to \Omega$
as follows: for $x,x'\in U(X)$,
$$
I^{\delta}_v(x,x')=
\begin{cases}
  U(v)(x) & \mbox{if } x=x' \\
  \bot & \mbox{otherwise}
\end{cases}
$$
If there is 
${\rm Eq}_{\delta}: {\rm Hom}_{\bf C}(X,\Omega)\to {\rm Hom}_{\bf C}(X\times X,\Omega)$
such that for any $v\in {\rm Hom}_{\bf C}(X,\Omega)$,
$${\rm Eq}_{\delta}(v)\in {\rm Hom}_{\bf C}(X\times X,\Omega) \mbox{ and } U({\rm Eq}_{\delta}(v))=I^{\delta}_v,$$
then the duality $T$-hyperdoctrine 
${\rm Hom}_{\bf C}(\mbox{-},\Omega):{\bf C}^{\rm op}\to {\bf Alg}(T)$ has equality $=$.
\end{prop}

In the case of comprehension, we make the following additional assumption
on the lifting of restricted maps that originally come from arrows in {\bf C}:
for any arrow $f:Y\to X$ in ${\bf C}$ and any $A\subset U(X)$,
if there is $X'\in {\bf C}$ with $U(X')=A$, then 
the restriction of $U(f)$ to $A$ lifts to an arrow in ${\bf C}$, i.e.,
there is an arrow $f':Y\to X'$ in ${\bf C}$ such that $U(f')$ is the restriction of $U(f)$ to $A$.
This actually holds in most concrete categories including 
the category of topological spaces and the category of algebras of a monad on {\bf Set}.

\begin{prop}\label{compre}
Consider a duality $T$-hyperdoctrine 
${\rm Hom}_{\bf C}(\mbox{-},\Omega):{\bf C}^{\rm op}\to {\bf Alg}(T)$,
its fibred category $\int {\rm Hom}_{\bf C}(\mbox{-},\Omega)$ 
derived via the Grothendieck construction, 
and the truth functor (see Definition \ref{compre})
$$\top:{\bf C}\to \int {\rm Hom}_{\bf C}(\mbox{-},\Omega).$$
If $U(\top_{{\rm Hom}(X,\Omega)})(x)=\top_{\Omega}$ 
for every $X\in {\bf C}$ and $x\in U(X)$,
and if there is a functor
$$Z:\int {\rm Hom}_{\bf C}(\mbox{-},\Omega)\to {\bf C}$$
such that the following hold:
\begin{itemize}
\item for $(X,v)\in \int {\rm Hom}_{\bf C}(\mbox{-},\Omega)$,
$U(Z(X,v)) = \{ x\in U(X) \ | \ U(v)(x)=\top_{\Omega} \}$;
\item for an arrow $(f,k)$ in $\int {\rm Hom}_{\bf C}(\mbox{-},\Omega)$,
$U(Z(f,k))=U(f)$,
\end{itemize}
then the duality $T$-hyperdoctrine 
${\rm Hom}_{\bf C}(\mbox{-},\Omega):{\bf C}^{\rm op}\to {\bf Alg}(T)$ 
has comprehension $\{\mbox{-}\}$
(the assumption intuitively means the correspondence 
$(X,v)\mapsto \{ x\in X \ | \ v(x)=\top_{\Omega} \}$ with $(f,k)\mapsto f$ 
over ${\bf Set}$ lifts to that over ${\bf C}$).
\end{prop}

All the assumptions of the propositions above are satisfied if ${\bf C}={\bf Set}$,
i.e., if we consider the dual adjunction between ${\bf Set}$ and ${\bf Alg}(T)$
induced by any $\Omega\in {\bf Alg}(T)$ as a dualising object.
And then the corresponding duality $T$-hyperdoctrine
${\rm Hom}_{\bf Set}(\mbox{-},\Omega):{\bf Set}^{\rm op}\to {\bf Alg}(T)$
turns out to be a model of higher-order logic over $T$.

\begin{thm}[Tarskian Models]\label{Tarski}
Let ${\bf C}={\bf Set}$ in a duality $T$-hyperdoctrine, i.e., consider
$${\rm Hom}_{\bf Set}(\mbox{-},\Omega):{\bf Set}^{\rm op}\to {\bf Alg}(T).$$
This ${\bf Set}$-${\bf Alg}$-duality $T$-hyperdoctrine is a higher-order $T$-hyperdoctrine.
\end{thm}

The most basic case is the powerset hyperdoctrine 
${\rm Hom}_{\bf Set}(\mbox{-},{\bf 2}):{\bf Set}^{\rm op}\to {\bf BA}$
where ${\bf BA}$ is the category of boolean algebras, and ${\bf 2}$ is the two-element algebra.
Interpretations in the powerset hyperdoctrine precisely captures the ordinary Tarski semantics
for first-order classical logic.

It is shown in Maruyama \cite{Mar} that such Tarskian hyperdoctrinal models yield 
sound and complete semantics for a wide variety of substructural logics as well as structured ones
(through choosing a monad $T$ or a variety of algebras in a suitable manner).

Objects in ${\bf Set}$ give domains of discourse for semantics,
and the dualising object $\Omega$ a set of truth values. In general, we need a class of different $\Omega$'s
to obtain completeness results, even though ${\bf 2}$ only suffices in the particular case of classical logic.
Such issues are discussed in Maruyama \cite{Mar}.


\section{Convex and Topological Geometric Logics}\label{geom}

In this section we consider applications of the theory above 
to topological geometric logic and convex geometric logic,
which illustrate what the theory means in concrete situations.
``Topological geometric logic" in our terms is usually called just ``geometric logic"
(it is the logic that is invariant under geometric morphisms of toposes).

Let us think of well-known dual adjunctions 
between topological spaces ${\bf Top}$ and frames ${\bf Frm}$,
and in particular its predicate functor 
$${\rm Hom}_{\bf Top}(\mbox{-},\Omega):{\bf Top}^{\rm op}\to {\bf Frm}$$
where it should be noted that not only the two-element frame ${\bf 2}$ but also 
any frame $\Omega$ induces a dual adjunction between ${\bf Top}$ and ${\bf Frm}$; 
this is a simple consequence of general duality theory
(any of duality theories \cite{Jo1,Marx,DT} works for this purpose).

The following is a consequence of Proposition \ref{exists} above.

\begin{cor}
The duality hyperdoctrine 
${\rm Hom}_{\bf Top}(\mbox{-},\Omega):{\bf Top}^{\rm op}\to {\bf Frm}$
has existential quantifier $\exists$.
In particular, the open set hyperdoctrine
${\rm Hom}_{\bf Top}(\mbox{-},{\bf 2}):{\bf Top}^{\rm op}\to {\bf Frm}$
has existential quantifier $\exists$. 
Thus, they give hyperdoctrine models of (topological) geometric logic.
\end{cor}

To exemplify the underlying idea of this,
let us consider the simplest case of the open set functor. 
It is then crucial to notice that $E^{\pi}_v$ in Proposition \ref{exists} 
gives us an open set by taking the inverse image of $1\in {\bf 2}$ under it.
This is true because any topology is closed under arbitrary unions.
Since a topology is not necessarily closed under arbitrary intersections,
the predicate functors above do not necessarily have universal quantifier.
Note that (topological) geometric logic does not have universal quantifier.

There are dual adjunctions between convex structures and Scott's continuous lattices 
(see Jacobs \cite{Jacdual} and Maruyama \cite{M-APAL,Marx}; 
the Jacobs duality for preframes can be recasted in terms of continuous lattices).
In the light of those dualities, 
we consider Scott's continuous lattices ${\bf ContLat}$ to represent pointfree convex structures,
just as frames represent pointfree topological spaces.

There are two concepts of abstract convex structures, and accordingly two kinds of dual adjunctions.
Let us denote by ${\bf Conv}$ the category of convexity spaces (for details, see van de Vel \cite{Vel}),
and by ${\bf Alg}({\mathcal D})$ the category of algebras of the distribution monad ${\mathcal D}$,
or equivalently barycentric algebras (for details, see Jacobs \cite{Jacdual}).
The following is a consequence of Proposition \ref{forall} above.

\begin{cor}
The ${\bf Conv}$-based duality hyperdoctrine 
${\rm Hom}_{\bf Conv}(\mbox{-},\Omega):{\bf Conv}^{\rm op}\to {\bf ContLat}$
has universal quantifier $\forall$.
The ${\bf Alg}({\mathcal D})$-based duality hyperdoctrine 
${\rm Hom}_{{\bf Alg}({\mathcal D})}(\mbox{-},\Omega):{\bf Alg}({\mathcal D})^{\rm op}\to {\bf ContLat}$
has universal quantifier $\forall$. 
\end{cor}

Thus, they give hyperdoctrine models of ``convex geometric logic",
which does not have existential quantifier, since in general
the set of convex subsets is not closed under arbitrary unions.

We can even apply the same idea to a dual adjunction 
between measurable spaces and $\sigma$-complete Boolean algebras
(see Maruyama \cite{Marxx,Marx}).

\section{Categorical Quantum Logic}\label{quant}

There are different conceptions of quantum logic and its algebras.
The lattice of projection operators on a Hilbert space 
is a standard algebra of quantum logic.
We can think of different categories encompassing those standard algebras of quantum logic,
including the category of orthomodular lattices, denoted ${\bf OML}$ 
and the category of effect algebras, denoted ${\bf EA}$.
The latter is more general than the former, and encompasses 
the algebra of effects of a Hilbert space 
as well as the algebra of projection operators.
Both ${\bf OML}$ and ${\bf EA}$ are algebraic categories, i.e., 
can be described as categories of algebras of monads on ${\bf Set}$.
Effect algebras only have negation and partial disjunction, 
and thus they are logically less expressive than orthomodular lattices.
In this section, we mainly work with ${\bf OML}$, and variants of it.

Fix a Hilbert space $H$, and let ${\rm P}(H)$ denote the lattice of projection operators on $H$.
We can see ${\rm P}(H)$ both as a set and as an algebra, 
and hence the set-algebra adjunction is available
(note that ${\rm Hom}_{\bf Set}(X,{\rm P}(H))$ is closed under 
the pointwise operations induced by the operations of ${\rm P}(H)$).
Let us consider the logic of the dual adjunction, i.e., regard
${\rm Hom}_{\bf Set}(\mbox{-},{\rm P}(H)): {\bf Set}^{\rm op} \to {\bf OML}$
as a $Q$-hyperdoctrine where $Q$ is the monad 
corresponding to the category ${\bf OML}$ of orthomodular lattices.
For the brevity of description, we drop the subscript  ``${\bf Set}$" 
of ``${\rm Hom}_{\bf Set}(\mbox{-},{\rm P}(H))$".

Now, Theorem \ref{Tarski} above tells us that
the set-based duality hyperdoctrine 
${\rm Hom}(\mbox{-},{\rm P}(H))$
forms a model of higher-order quantum logic:

\begin{cor}
The set-based duality hyperdoctrine ${\rm Hom}(\mbox{-},{\rm P}(H))$ is a higher-order $Q$-hyperdoctrine (or $Q$-tripos).
More generally, ${\rm Hom}(\mbox{-},\Omega)$ for any $\Omega\in {\bf OML}$ is a higher-order $Q$-hyperdoctrine.
\end{cor}

In the following we look at the above type of hyperdoctrines from two different perspectives.

\subsection{The Tripos-to-Topos Construction and Quantum Set Theory}

Given a frame $\Omega$, the set-based duality hyperdoctrine 
${\rm Hom}(\mbox{-},\Omega):{\bf Set}^{\rm op}\to {\bf Frm}$
yields via the tripos-to-topos construction
the Higgs topos of $\Omega$-valued sets, or equivalently the sheaf topos on $\Omega$,
or equivalently the topos of sets in the $\Omega$-valued model of set theory (aka. Heyting-valued models; 
see, e.g., Bell \cite{Bell}).

Let us think of a quantum analogue of this.
The tripos-to-topos construction in the present context can be defined 
in the same way as in Maruyama \cite[Definition 14]{Mar}, 
as the category 
${\bf T}({\rm Hom}(\mbox{-},{\rm P}(H)))$ 
of partial equivalence relations in the internal logic of a given 
${\rm Hom}(\mbox{-},{\rm P}(H))$.
Note that we only need deductive relations (i.e., partial orders on fibres), 
conjunction, and existential quantifier
when defining the tripos-to-topos construction;
they indeed exist in ${\rm Hom}(\mbox{-},{\rm P}(H))$.

Now our question is how ${\bf T}({\rm Hom}(\mbox{-},{\rm P}(H)))$ 
compares to the known concept of Takeuti-Ozawa's quantum set theory, 
to be precise the ${\rm P}(H)$-valued model of set theory, which is defined as follows:
for each ordinal $\alpha$, define via the transfinite recursrion
$V_{\alpha}=\{ u \ | \ u:D\to {\rm P}(H) \mbox{ and } D\subset \bigcup_{\beta \leq \alpha} V_{\beta}\}$
and then let $V=\bigcup_{\alpha\in {\rm Ord}}V_{\alpha}$ where ${\rm Ord}$ is the class of all ordinals.
We denote by ${\bf Set}^{{\rm P}(H)}$ the category of sets in this model of set theory.
We then have the following proposition.

\begin{prop}
${\bf T}({\rm Hom}(\mbox{-},{\rm P}(H)))$ embeds into ${\bf Set}^{{\rm P}(H)}.$
\end{prop}

We next think of completeness wrt. proof-theoretic calculus, 
which has so far been lacking in categorical quantum logic with quantifiers
(that without quantifiers has already been developed).

\subsection{Faggian-Sambin's Calculus over Cartesian/Monoidal Type Theory}

Let ${\bf QL}$ denote the category of algebras of Faggian-Sambin's propositional quantum logic ${\rm FS}$;
algebraisation of logic is automatic via the well-known methods of Abstract Algebraic Logic.
For syntactic details, we refer to Faggian-Sambin \cite{FS}, due to space limitations.
${\rm FS}$ can be quantified in the same way as Sambin's Basic Logic.
The quantified ${\rm FS}$ can then be typed in the same manner as 
typed intuitionistic logic in Pitts \cite{Pit}, or typed Full Lambek calculus in Maruyama \cite{Mar}.
We denote by ${\rm TFS}^q$ the resulting typed quantum sequent calculus.

\begin{thm}
The class of all ${\bf QL}$-hyperdoctrines $P:{\bf C}^{\rm op}\to {\bf QL}$ gives sound and complete semantics for ${\rm TFS}^q$.
\end{thm}

We can refine the theorem above into the following by focusing upon set-based duality hyperdoctrines in Theorem \ref{Tarski}.

\begin{thm}
The class of all set-based duality ${\bf QL}$-hyperdoctrines 
${\rm Hom}(\mbox{-},\Omega):{\bf Set}^{\rm op}\to {\bf QL}$
where $\Omega\in {\bf QL}$ gives sound and complete semantics for ${\rm TFS}^q$.
\end{thm}

We may even replace the cartesian type theory of the logic by the monoidal one,
in the same way as Ambler \cite{Amb} considers logic over monoidal type theory.
This is a merit of the hyperdoctrine approach, in which logic and type theory are separated, 
and can be chosen independently of each other.
That is, we choose Faggian-Sambin's quantum calculus for the logic part, 
and Ambler's linear type theory for the type theory part, 
which amounts to SMCC (symmetric monoidal closed categories).
Accordingly, the base category of a hyperdoctrine is taken to be an SMCC with finite products;
note that we still keep cartesian products for the purpose of defining quantifiers.
Let ${\rm LFS}^q$ denote the linearly typed quantum sequent calculus.

\begin{thm}
The class of all ${\bf QL}$-hyperdoctrines $P:{\bf C}^{\rm op}\to {\bf QL}$ over SMCC ${\bf C}$ with products
gives sound and complete semantics for ${\rm LFS}^q$.
\end{thm}

In the Hilbert hyperdoctrine ${\rm KSub}:{\bf Hilb}^{\rm op}\to {\bf QL}$,
tensor $\otimes$ maps two projections $P\in {\rm KSub}(X)$ and $Q\in {\rm KSub}(Y)$ into 
$P\otimes Q\in {\rm KSub}(X\otimes Y)$, i.e., it functions as translation between different fibres.
We consider that dagger-SMCC-based quantum-logic-valued hyperdoctrines 
enriched with a structure to express this tensor translation between fibres
give a synthesis of Birkhoff-von Neumann's quantum logic and Abramsky-Coecke's categorical quantum mechanics.






\end{document}